\documentstyle[12pt]{article}  
\begin{document}  \bibliographystyle{unsrt}

\begin{center}
{\Large \bf Lorentz Boosts as Squeeze Transformations and the Parton
Picture}~\footnote{presented at the Conference on Fundamental Interaction
of Elementary Particles, An Annual Meeting of Particle and Nuclear Physics
Division of the Russian Academy of Sciences.}
\\[6mm]

Y. S. Kim~\footnote{email: kim@umdhep.umd.edu}   \\
{\it Department of Physics, University of Maryland, \\College Park,
Maryland 20742, U.S.A.}
\end{center}

\vspace{4mm}

\begin{abstract}
It was shown by Gribov, Ioffe, Pomeranchuk in 1966 and by Ioffe in
1969 that a space-time picture is needed for the Lorentz deformation
of hadronic interaction region.  It is shown that this deformation
is a squeeze transformation.  It is shown also that Feynman's parton
picture emerges as a consequence of Lorentz-squeezed hadrons in
the quark model.

\end{abstract}

\section{Introduction}\label{intro}
According to special relativity, the longitudinal length of a moving
object becomes contracted while the transverse components remain
invariant.  There is therefore a tendency to assume that hadrons from
an accelerator look like ``pancakes'' with a contracted longitudinal
dimension~\cite{bj69}.  Yes, an extended hadron should have the
three-dimensional rotational symmetry when it is on the
table~\cite{wig39}.  If it moves with a speed close to that of light,
it should look different.  For static or slow hadrons, we use the
quark model to understand what we observe in laboratories.  For
fast-moving hadrons, we use the parton picture to interpret the
experimental data.  In both the quark and parton models, quarks or
partons interact directly with the external signal.  In the quark model,
we add those interaction amplitudes before calculating the cross section.
On the other hand, in the parton model, we calculate the cross section
for each parton before summing up the cross sections for all the partons.
Does this mean that the Lorentz boost destroys the superposition
principle?

In order to answer this question, let us examine the ``pancake effect''
more carefully.  This effort was started by Gribov, Ioffe, and
Pomeranchuk in 1966~\cite{gribov66}.  According to the old-fashioned
picture of Lorentz pancakes, only the longitudinal component becomes
contracted.   In their 1966 paper~\cite{gribov66}, Gribov {\it et al.}
showed that combinations of space and time variables are needed
in measuring the dimension of the interaction region as well as the
interaction time.  They showed further that the interaction time is
proportional to the contracted component of the space-time variables
which are known today as the light-cone variables.  In his 1969
paper~\cite{ioffe69}, Ioffe essentially completed the Lorentz-squeeze
picture of the interaction region.  This squeeze picture was shown
to be convenient in explaining the high-energy data by Drell and
Yan~\cite{drell71}.  This picture is illustrated in one of the figures
in the Drell-Yan paper.  Figure~1 of the present paper reproduces the
Lorentz squeeze property formulated by these authors.

For a hadron with its space-time extension, the interaction region is
essentially the region in which the quarks are distributed.  Thus, the
problem is reduced to the study of space-time distribution of the quarks
inside the hadron.  This means that we have to learn how to boost wave
functions in quantum mechanics.  This problem is then reduced to that
of constructing Lorentz-covariant wave functions.  While this is not a
trivial problem, it is possible to construct a model based within the
framework of Wigner's little groups which dictate the internal
space-time symmetries of relativistic particles~\cite{wig39}.

If we are to construct covariant wave functions, they should possess the
symmetry of the little groups.  After constructing such a set of wave
functions, we should be able to take both the low-speed and high-speed
limits of the wave functions to generate the quark and parton models
respectively.  Indeed, there is in the literature a formalism of
covariant bound-state wave functions which can be Lorentz-boosted. It is
called the covariant harmonic oscillator formalism~\cite{knp86,dir45}.
The formalism meets the following three basic requirements.

\begin{itemize}

\item[1).]  The formalism is consistent with the established physical
principles including the uncertainty principle in quantum mechanics
and the transformation laws of special relativity~\cite{knp86}.

\item[2).] The formalism is consistent with the basic hadronic
features observed in high-energy laboratories, including hadronic mass
spectra, the proton form factor, and the parton phenomena~\cite{knp86}.

\item[3).] The formalism constitutes a representation of the
Poincar\'e group for relativistic extended hadrons~\cite{kno79a}, and
a representation of Wigner's little group.
\end{itemize}

\noindent In addition, this oscillator system provides the mathematical
basis for a certain set of coherent photon states commonly known as the
squeezed state of light~\cite{knp91}.  Through this formalism, we are
able to see clearly that Lorentz boosts are squeeze transformations.

In this paper, we use this covariant oscillator formalism to see that
the quark model and the parton models are two different manifestations
of one covariant formalism.  We shall see how the parton picture
emerges from the Lorentz-squeezed hadronic wave function.  This squeeze
effect will also explain why the partons appear as incoherent particles,
within the present framework of quantum mechanics based on the
superposition principle.

Since we are going to use the language of little groups in this
paper, we give in Sec.~\ref{little} a historical review of Wigner's
little groups.  In Sec.~\ref{boost}, we use the light-cone coordinate
system to show that Lorentz boosts are squeeze transformations.
Section~\ref{cov} contains an outline of the covariant oscillator
formalism which will exhibit the squeeze property of Lorentz boosts
in quantum mechanics.  Finally, in Sec.~\ref{parton}, it is shown that
the covariant oscillator wave function gives a static wave function for
the hadron at rest and the parton distribution function for the hadron
moving with a speed very close to the speed of light.  It is shown also
that the time interval for the quark to interact with the external signal
becomes contracted while the interval for the quark-quark interaction
becomes dilated.

\section{Wigner's Little Groups}\label{little}
From the principles of special relativity, Einstein derived the relation
$E = mc^{2}$ in 1905.  This formula unifies the momentum-energy relations
for both massive and massless particles, which are $E = p^{2}/2m$ and
$E = cp$ respectively.  In addition to the energy-momentum variables,
relativistic particles have internal space-time degrees of freedom.
A massive particle at rest has three rotational degrees of freedom, and
they appear in the real world as the spin of the particle.  Massless
particles have only one rotational degree of freedom which appears as
the helicity in the real world.  In addition, they have gauge degrees of
freedom which are not shared by massive particles.  Why are these two
symmetries different from each other?  Is it possible to unify the
symmetries for both cases as Einstein did for the energy-momentum
relation?  This problem is summarized in Fig. 2.

In his 1939 paper~\cite{wig39}, Wigner took the first step toward the
resolution of this problem.  He observed that the internal space-time
symmetries of relativistic particles are dictated by their respective
little groups.  The little group is the maximal subgroup of the
Lorentz group which leaves the four-momentum of the particle invariant.
He showed that the little groups for massive and massless particles
are isomorphic to $O(3)$ (three-dimensional rotation group) and
$E(2)$ (two-dimensional Euclidean group) respectively.  Wigner's 1939
paper indeed gives a covariant picture massive particles with spins, and
connects the helicity of massless particle with the rotational degree
of freedom in the group $E(2)$.  This paper also gives many homework
problems, including the following four pressing problems in particle
physics.

\begin{itemize}

\item[1).]  Like the three-dimensional rotation group, $E(2)$ is a
three-parameter group.  It contains two translational degrees of freedom
in addition to the rotation.  What physics is associated with the
translational-like degrees of freedom for the case of the $E(2)$-like
little group?

\item[2).]  As is shown by Inonu and Wigner~\cite{inonu53}, the rotation
group $O(3)$ can be contracted to $E(2)$.  Does this mean that the
O(3)-like little group can become the $E(2)$-like little group in a
certain limit?

\item[3).]  It is possible to interpret the Dirac equation in terms of
Wigner's representation theory~\cite{barg48}.  Then, why is it not
possible to find a place for Maxwell's equations in the same theory?

\item[4).]  The proton was found to have a finite space-time extension
in 1955~\cite{hofsta55}, and the quark model has been established in
1964~\cite{gell64}.  The concept of relativistic extended particles has
now been firmly established.  Is it then possible to construct a
representation of the Poincar\'e group for particles with space-time
extensions?
\end{itemize}

Indeed, there are many papers written in the literature on the
above-mentioned problems~\cite{knp86,kim95mex}, and the present
situation is summarized in Fig.~2.  In this report, we are interested
only in the fourth question.  It is about whether Wigner's little
groups are applicable to high-energy hadrons fresh from particle
accelerators.  The question is whether it is possible to construct a
representation of the little group for hadrons which are believed
to be quantum bound states of quarks~\cite{knp86,fkr71}.  This
representation should describe Lorentz-boosted hadrons.  The next
question is whether those boosted wave functions generate Feynman's
parton picture~\cite{fey69} in the large-momentum limit.

Within the framework of Wigner's little groups, the ultimate question
is whether the quark model and the parton model cam be framed into the
$O(3)$-like little group for massive particles and the $E(2)$-like
little group for massless particles~\cite{kim89}.  This mathematical
question is beyond the scope of the present paper.

\section{Lorentz Boosts as Squeeze Transformations}\label{boost}
The boost matrix for the longitudinal and time-like variables takes the
form
\begin{equation}\label{trans1}
B(\eta) = \pmatrix{\cosh\eta  & \sinh\eta \cr
\sinh\eta & \cosh\eta },
\end{equation}
applicable to the column matrix of $(z, t)$, with $tanh\eta = \beta$
where $\beta$ is the velocity parameter of the hadron.  In 1949, Dirac
chose the coordinate variables~\cite{dir49}
\begin{equation}\label{uv}
u = (z + t)/\sqrt{2}, \qquad v = (z - t)/\sqrt{2} ,
\end{equation}
in order to simplify the formula for Lorentz boosts.  The boost matrix
applicable to the column vector $(u, v)$ now becomes diagonal and takes
the form
\begin{equation}\label{trans2}
B(\eta) = \pmatrix{\exp{(\eta)}  & 0 \cr 0 & \exp{(-\eta)} } .
\end{equation}
The $u$ and $v$ variables are called the light-cone variables.
Under this transformation, the $u$ and $v$ variables become expanded and
contracted by the factors $\exp{\eta}$ and $\exp(-\eta)$ respectively.
The product $uv$ remains invariant.  From Fig. 1, it is quite clear that
the Lorentz boost is an area-preserving ``squeeze'' transformation.

The transformation matrix of Eq.(\ref{trans1}) is applicable also to the
momentum-energy column matrix $(P, E)$, where $P$ and $E$ are the
longitudinal momentum and the total energy respectively.  As for the
light-cone variables
\begin{equation}
P_{+} = (P + E)/\sqrt{2} , \qquad P = (P - E)/\sqrt{2} ,
\end{equation}
the transformation matrix is Eq.(\ref{trans2}).  This is also a squeeze
transformation.

The word ``squeeze'' is commonly used these days in quantum optics for
a certain class of two-photon coherent states~\cite{knp91}, but the
concept is squeeze transformations is applicable to many different
branches of physics, including the Lorentz boost so fundamental in
high-energy physics.

\section{Covariant Harmonic Oscillators}\label{cov}
If we construct a representation of the Lorentz group using normalizable 
harmonic oscillator wave functions, the result is the covariant harmonic 
oscillator formalism~\cite{knp86,dir45}.  The formalism constitutes a
representation of Wigner's $O(3)$-like little group for a massive
particle with internal space-time structure.  This oscillator formalism
has been shown to be effective in explaining the basic phenomenological
features of relativistic extended hadrons observed in high-energy
laboratories.  In particular, the formalism shows that the quark model
and Feynman's parton picture are two different manifestations of one
covariant entity~\cite{knp86,kim89}.

The covariant harmonic oscillator formalism has been discussed
exhaustively in the literature, and it is not necessary to give another
full-fledged treatment in the present paper.  We shall discuss here
only the squeeze property of the oscillator wave functions.
Let us consider a bound state of two particles.  For convenience, we
shall call the bound state the hadron, and call its constituents quarks.
Then there is a Bohr-like radius measuring the space-like separation
between the quarks.  There is also a time-like separation between the
quarks, and this variable becomes mixed with the longitudinal spatial
separation as the hadron moves with a relativistic speed.  There are
no quantum excitations along the time-like direction.  On the other
hand, there is the time-energy uncertainty relation which allows quantum
transitions.  It is possible to accommodate these aspect within the
framework of the present form of quantum mechanics.  The uncertainty
relation between the time and energy variables is the c-number
relation~\cite{dir27}, which does not allow excitations along the
time-like coordinate.  We shall see that the covariant harmonic
oscillator formalism accommodates this narrow window in the present form
of quantum mechanics.

Let us consider now a hadron consisting of two quarks.  If the space-time
position of two quarks are specified by $x_{a}$ and $x_{b}$ respectively, 
the system can be described by the variables 
\begin{equation}
X = (x_{a} + x_{b})/2 , \qquad x = (x_{a} - x_{b})/2\sqrt{2} .
\end{equation}
The four-vector $X$ specifies where the hadron is located in space and
time, while the variable $x$ measures the space-time separation between
the quarks.  In the convention of Feynman {\it et al.}~\cite{fkr71}, the
internal motion of the quarks bound by a harmonic oscillator potential
of unit strength can be described by the Lorentz-invariant equation
\begin{equation}
{1\over 2}\left\{x^{2}_{\mu } -
{\partial ^{2} \over \partial x_{\mu }^{2}} \right\} \psi (x)
= \lambda \psi (x) .
\end{equation}
We use here the space-favored metric: $x^{\mu } = (x, y, z, t)$.

It is possible to construct a representation of the Poincar\'e group
from the solutions of the above differential equation~\cite{knp86}.  If
the hadron is at rest, the solution should take the form
\begin{equation}
\psi (x,y,z,t) = \psi (x,y,z) \left({1\over \pi }\right)^{1/4} 
\exp{\left(-t^{2}/2 \right)} ,
\end{equation}
where $\psi (x,y,z)$ is the wave function for the three-dimensional 
oscillator with appropriate angular momentum quantum numbers.  Indeed,
the above wave function constitutes a representation of Wigner's
$O(3)$-like little group for a massive particle~\cite{knp86}.  In the
above expression, there are no time-like excitations, and this is
consistent with what we see in the real world.  It was Dirac who noted
first this space-time asymmetry in quantum mechanics~\cite{dir27}.
However, this asymmetry is quite consistent with the $O(3)$ symmetry of
the little group for hadrons.  Figure 3 illustrates the uncertainty
relations along the space-like and time-like directions.

Since the three-dimensional oscillator differential equation is
separable in both spherical and Cartesian coordinate systems,
$\psi (x,y,z)$ consists of Hermite polynomials of $x, y$, and $z$.  If
the Lorentz boost is made along the $z$ direction, the $x$ and $y$
coordinates are not affected, and can be dropped from the wave function.
The wave function of interest can be written as
\begin{equation}
\psi^{n}(z,t) = \left({1\over \pi }\right)^{1/4}
\exp{\left(-t^{2}/2\right)}\psi_{n}(z) ,
\end{equation}
with
\begin{equation}
\psi ^{n}(z) = \left({1 \over \pi n!2^{n}} \right)^{1/2} H_{n}(z) 
\exp (-z^{2}/2) ,
\end{equation}
where $\psi ^{n}(z)$ is for the $n$-th excited oscillator state.  
The full wave function $\psi ^{n}(z,t)$ is
\begin{equation}\label{wf1}
\psi ^{n}_{0}(z,t) = \left({1\over \pi n! 2^{n}}\right)^{1/2} H_{n}(z) 
\exp \left\{-{1\over 2}\left(z^{2} + t^{2} \right) \right\} .
\end{equation}
The subscript 0 means that the wave function is for the hadron at rest.
The above expression is not Lorentz-invariant, and its localization
undergoes a Lorentz squeeze as the hadron moves along the $z$
direction~\cite{knp86}.  This is a Lorentz-covariant expression!

Let us write the above wave functions in terms of the light-cone
variables defined in Eq.(\ref{uv}).  The wave function of Eq.(\ref{wf1})
can be written as
\begin{equation}\label{wf2}
\psi^{n}_{o}(z,t) = \psi^{n}_{0}(z,t)
= \left({1 \over \pi n!2^{n}} \right)^{1/2} H_{n}\left((u + v)/\sqrt{2}
\right) \exp \left\{-{1\over 2} (u^{2} + v^{2}) \right\} . 
\end{equation}
If the system is boosted, the wave function becomes
\begin{equation}\label{wf3}
\psi ^{n}_{\eta }(z,t) = \left({1 \over \pi n!2^{n}} \right)^{1/2} 
H_{n} \left((e^{-\eta }u + e^{\eta }v)/\sqrt{2} \right) 
\times \exp \left\{-{1\over 2}\left(e^{-2\eta }u^{2} + 
e^{2\eta }v^{2}\right)\right\} .
\end{equation}
Indeed, in the light-cone coordinate system, the Lorentz-boosted wave
function takes a very simple form.

In both Eqs. (\ref{wf2}) and (\ref{wf3}), the localization property
of the wave function in the $u v$ plane is determined by the Gaussian
factor, and it is sufficient to study the ground state only for the
essential feature of the boundary condition.  The wave functions in
Eq.(\ref{wf2}) and Eq.(\ref{wf3}) then respectively become
\begin{equation}\label{wf4}
\psi _{0}(z,t) = \left({1 \over \pi} \right)^{1/2} 
\exp \left\{-{1\over 2} (u^{2} + v^{2}) \right\} . 
\end{equation}
If the system is boosted, the wave function becomes
\begin{equation}\label{wf5}
\psi _{\eta }(z,t) = \left({1 \over \pi }\right)^{1/2} 
\exp \left\{-{1\over 2}\left(e^{-2\eta }u^{2} + 
e^{2\eta }v^{2}\right)\right\} .
\end{equation}
The transition from Eq.(\ref{wf4}) to Eq.(\ref{wf5})
is a squeeze transformation.  The wave function of Eq.(\ref{wf4}) is
distributed within a circular region in the $u v$ plane, and thus in
the $z t$ plane.  On the other hand, the wave function of Eq.(\ref{wf5})
is distributed in an elliptic region.  This ellipse is a ``squeezed''
circle with the same area as the circle, as is illustrated in Fig. 4.
The Lorentz boost squeezes the oscillator wave function.

\section{Feynman's Parton Picture}\label{parton}

It is safe to believe that hadrons are quantum bound states of quarks
having localized probability distribution.  As in all bound-state cases,
this localization condition is responsible for the existence of discrete
mass spectra.  The most convincing evidence for this bound-state picture
is the hadronic mass spectra which are observed in high-energy
laboratories~\cite{knp86,fkr71}.  However, this picture of bound states
is applicable only to observers in the Lorentz frame in which the hadron
is at rest.  How would the hadrons appear to observers in other Lorentz
frames?  More specifically, can we use the picture of Lorentz-squeezed
hadrons discussed in Sec.~\ref{cov}.

It was Hofstadter's experiment which showed that the proton charge is
spread out.  In this experiment, an electron emits a virtual photon,
which then interacts with the proton.  If the proton consists of quarks
distributed within a finite space-time region, the virtual photon will
interact with quarks which carry fractional charges.  The scattering
amplitude will depend on the way in which quarks are distributed within
the proton.  The portion of the scattering amplitude which describes the
interaction between the virtual photon and the proton is called the
form factor.

Although there have been many attempts to explain the behavior of form
factors within the framework of quantum field theory, it is quite natural
to expect that the wave function in the quark model will determine the
charge distribution.  In high-energy experiments, we are dealing with the
situation in which the momentum transfer in the scattering process is
large.  Indeed, the Lorentz-squeezed wave functions lead to the correct
behavior of the hadronic form factor for large values of the momentum
transfer~\cite{fuji70}.

While the form factor is the quantity which can be extracted from the 
elastic scattering, it is important to realize that in high-energy 
processes, many particles are produced in the final state.  They are
called inelastic processes.  While the elastic process is described by
the total energy and momentum transfer in the center-of-mass coordinate
system, there is, in addition, the energy transfer in inelastic
scattering.  Therefore, we would expect that the scattering cross
section would depend on the energy, momentum transfer, and energy
transfer.  However, one prominent feature in inelastic scattering is
that the cross section remains nearly constant for a fixed value of
the momentum-transfer/energy-transfer ratio.  This phenomenon is called
``scaling''~\cite{bj69}.

In order to explain the scaling behavior in inelastic scattering,
Feynman in 1969 observed that a fast-moving hadron can be regarded as
a collection of many ``partons'' whose properties do not appear to be
identical with those of the quarks~\cite{fey69}.  For example, the number
of quarks inside a static proton is three, while the number of partons in
a rapidly moving proton appears to be infinite.  The question then is how
the proton looking like a bound state of quarks to one observer can
appear different to an observer in a different Lorentz frame?  Feynman
formulated his parton picture based on the following observations.
\begin{itemize}
\item[1).] The picture is valid only for hadrons moving with velocity close
       to that of light.

\item[2).] The interaction time between the quarks becomes dilated, and
        partons behave as free independent particles.

\item[3).] The momentum distribution of partons becomes widespread as the
       hadron moves fast.

\item[4).] The number of partons seems to be infinite or much larger than
       that of quarks.

\end{itemize}

\noindent Because the hadron is believed to be a bound state of two or three 
quarks, each of the above phenomena appears as a paradox, particularly 2) and
3) together.  We would like to resolve this paradox using the covariant
harmonic oscillator formalism. 

For this purpose, we need a momentum-energy wave function.  If the quarks 
have the four-momenta $p_{a}$ and $p_{b}$, we can construct two independent 
four-momentum variables~\cite{fkr71}
\begin{equation}
P = p_{a} + p_{b} , \qquad q = \sqrt{2}(p_{a} - p_{b}) .
\end{equation}
The four-momentum $P$ is the total four-momentum and is thus the hadronic 
four-momentum.  $q$ measures the four-momentum separation between the quarks.

We expect to get the momentum-energy wave function by taking the Fourier 
transformation of Eq.(\ref{wf5}):
\begin{equation}\label{fourier}
\phi_{\eta }(q_{z},q_{0}) = \left({1 \over 2\pi }\right) 
\int \psi_{\eta}(z, t) \exp{\left\{-i(q_{z}z - q_{0}t)\right\}} dx dt .
\end{equation}
Let us now define the momentum-energy variables in the light-cone coordinate 
system as 
\begin{equation}\label{conju}
q_{u} = (q_{0} - q_{z})/\sqrt{2} ,  \qquad
q_{v} = (q_{0} + q_{z})/\sqrt{2} .
\end{equation}
In terms of these variables, the Fourier transformation of
Eq.(\ref{fourier}) can be written as
\begin{equation}\label{fourier2}
\phi_{\eta }(q_{z},q_{0}) = \left({1 \over 2\pi }\right) 
\int \psi_{\eta}(z, t) \exp{\left\{-i(q_{u} u + q_{v} v)\right\}} du dv .
\end{equation}
The resulting momentum-energy wave function is 
\begin{equation}\label{phi}
\phi_{\eta }(q_{z},q_{0}) = \left({1 \over \pi }\right)^{1/2} 
\exp\left\{-{1\over 2}\left(e^{-2\eta}q_{u}^{2} + 
e^{2\eta}q_{v}^{2}\right)\right\} .
\end{equation}
Because we are using here the harmonic oscillator, the mathematical form 
of the above momentum-energy wave function is identical to that of the 
space-time wave function.  The Lorentz squeeze properties of these wave 
functions are also the same, as is indicated in Fig. 5.

When the hadron is at rest with $\eta = 0$, both wave functions behave
like those for the static bound state of quarks.  As $\eta$ increases,
the wave functions become continuously squeezed until they become
concentrated along their respective positive light-cone axes.  Let us
look at the z-axis projection of the space-time wave function.  Indeed,
the width of the quark distribution increases as the hadronic speed
approaches that of the speed of light.  The position of each quark
appears widespread to the observer in the laboratory frame, and the
quarks appear like free particles.

Furthermore, interaction time of the quarks among themselves become
dilated.  Because the wave function becomes wide-spread, the distance
between one end of the harmonic oscillator well and the other end
increases as is indicated in Fig. 4.  This effect, first noted by
Feynman~\cite{fey69}, is universally observed in high-energy hadronic
experiments.  Let us look at the time ratio more carefully.  The
period of oscillation increases like $e^{\eta}$ as was predicted by
Feynman~\cite{fey69}.

On the other hand, the quark's interaction time with the external
signal decreases as $e^{-\eta}$ as was predicted by Gribov
{\it et al.}~\cite{gribov66}.  In the picture of the Lorentz
squeezed hadron given in Fig. 4, the hadron moves along the $u$
(positive light-cone) axis, while the external signal moves in the
direction opposite to the hadronic momentum, which corresponds to
the $v$ (negative light-cone) axis.  This time interval is proportional
to the minor axis of the ellipse given in Fig. 4.

If we use $T_{ext}$ and $T_{osc}$ for the quark's interaction time with
external signal and the interaction time among the quarks, their ratio
becomes
\begin{equation}\label{ratio}
{T_{ext} \over T_{osc}} = \frac{\exp(-\eta)}{\exp(\eta)} = exp(-2\eta) .
\end{equation}
The ratio of the interaction time to the oscillator period becomes
$e^{-2\eta}$.  The energy of each proton coming out of the Fermilab
accelerator is $900 GeV$.  This leads the ratio to $10^{-6}$.  This is
indeed a small number.  The external signal is not able to sense the
interaction of the quarks among themselves inside the hadron.  Thus, the
quarks are free particles for the external signal.  This is the cause of
incoherence in the parton interaction amplitudes.

The momentum-energy wave function is just like the space-time wave
function in the oscillator formalism.  The longitudinal momentum
distribution becomes wide-spread as the hadronic speed approaches the
velocity of light.  This is in contradiction with our expectation from
nonrelativistic quantum mechanics that the width of the momentum
distribution is inversely proportional to that of the position wave
function.  Our expectation is that if the quarks are free, they must
have their sharply defined momenta, not a wide-spread distribution.
This apparent contradiction presents to us the following two fundamental
questions:

\begin{itemize}

\item[1).]  If both the spatial and momentum distributions become
      widespread as the hadron moves, and if we insist on Heisenberg's
      uncertainty relation, is Planck's constant dependent on the
      hadronic velocity?

\item[2).]  Is this apparent contradiction related to another apparent
      contradiction that the number of partons is infinite while there 
      are only two or three quarks inside the hadron?
\end{itemize}

The answer to the first question is ``No'', and that for the second
question is ``Yes''.  Let us answer the first question which is related
to the Lorentz invariance of Planck's constant.  If we take the product
of the width of the longitudinal momentum distribution and that of the
spatial distribution, we end up with the relation
\begin{equation}
<z^{2}><q_{z}^{2}> = (1/4)[\cosh(2\eta)]^{2}  .  
\end{equation}
The right-hand side increases as the velocity parameter increases.  This 
could lead us to an erroneous conclusion that Planck's constant becomes 
dependent on velocity.  This is not correct, because the longitudinal 
momentum variable $q_{z}$ is no longer conjugate to the longitudinal
position variable when the hadron moves.

In order to maintain the Lorentz-invariance of the uncertainty product,
we have to work with a conjugate pair of variables whose product does
not depend on the velocity parameter.  Let us go back to Eq.(\ref{conju})
and Eq.(\ref{fourier2}).  It is quite clear that the light-cone variable
$u$ and $v$ are conjugate to $q_{u}$ and $q_{v}$ respectively.  It is
also clear that the distribution along the $q_{u}$ axis shrinks as the
$u$-axis distribution expands.  The exact calculation leads to
\begin{equation}
<u^{2}><q_{u}^{2}> = 1/4 , \qquad  <v^{2}><q_{v}^{2}> = 1/4  .    
\end{equation}
Planck's constant is indeed Lorentz-invariant.  

Let us next resolve the puzzle of why the number of partons appears to
be infinite while there are only a finite number of quarks inside the
hadron.  As the hadronic speed approaches the speed of light, both the
$x$ and $q$ distributions become concentrated along the positive
light-cone axis.  This means that the quarks also move with velocity
very close to that of light. Quarks in this case behave like massless
particles.

We then know from statistical mechanics that the number of massless
particles  is not a conserved quantity.  For instance, in black-body
radiation, free light-like particles have a widespread momentum
distribution.  However, this does not contradict the known principles
of quantum mechanics, because the massless photons can be divided into
infinitely many massless particles with a continuous momentum
distribution.

Likewise, in the parton picture, massless free quarks have a wide-spread 
momentum distribution.  They can appear as a distribution of an 
infinite number of free particles.  These free massless particles are
the partons.  It is possible to measure this distribution in high-energy
laboratories, and it is also possible to calculate it using the covariant
harmonic oscillator formalism.  We are thus forced to compare these two 
results~\cite{hussar81}.  Figure 6 shows the result.

\section*{Concluding Remarks}
This is largely a review paper, but it contains the following new
observation.  Let us go to the time ratio given in Eq.(\ref{ratio}).
It is a product of two identical numbers.  The factor given by Feynman's
time dilation effect is $e^{-\eta}$.  The ratio given by the time
contraction effect of Gribov {\it et al.}  is also $e^{-\eta}$.  Thus
the combined effect is $e^{-2\eta}$.  This combined effect makes the
parton amplitudes to lose coherence even at moderate hadronic speed.

Another noteworthy point is that Wigner's little group is not only
an abstract concept, but also serves as a computational tool in
high-energy physics.  The covariant harmonic oscillator is one of the
tools derivable from the concept of little groups.  It is interesting
to note that the covariant oscillator formalism gives both Feynman's
time dilation and the time contraction of Gribov {\it et al}.

\section*{Acknowledgments}
The author is grateful to Prof. B. L. Ioffe for inviting him to
the Conference on Fundamental Interaction of Elementary Particles held
in October of 1995.  This Conference was one of the annual meetings of
the Division of Particle and Nuclear Physics of the Russian Academy of
Sciences, but this 1995 meeting had a special significance.  It was
held in commemoration of the 50th anniversary of the Institute of
Theoretical and Experimental Physics.  The author is indeed gratified
to have an opportunity to discuss one of the important contributions
made by the three distinguished members of this Institute.  He is
grateful also to many of his Russian colleagues, particularly A. S.
Chirkin, V. A. Isakov, G. Kotel'nikov, M. Man'ko and V. I. Man'ko,
for extending warm hospitality to him while in Moscow.

\newpage

\section*{Figure Captions}

\hspace{5mm} FIG. 1.
Space-time picture of the Lorentz boost.  The invariant quantiy
$(z^{2} - t^{2})$ can be written as $(z + t)(z - t)$.  This is
proportional to the product of the light-cone variables
$u = (z + t)/\sqrt{2}$ and $(z - t)/\sqrt{2}$.  The most appropriate
name for this area-preserving deformation is SQUEEZE.

FIG. 2.
Further implications of Einstein's $E = mc^{2}$.  Massive and
massless particles have different energy-momentum relations.  Einstein's 
special relativity gives one relation for both.  Winger's little group 
unifies the internal space-time symmetries for massive and massless
particles which are locally isomorphic to $O(3)$ and $E(2)$ respectively.
It is a great challenge for us to find another unification.  In this
note, we present a unified picture of the quark and parton models which
are applicable to slow and ultra-fast hadrons respectively.

FIG. 3.
Quantum mechanics with the c-number time-energy uncertainty relation. The
present form of quantum mechanics allows quantum excitations along the
space-like directions, but does not allow excitations along the time-like
direction even though there is an uncertainty relation between the time
and energy variables.

FIG. 4.
Relativistic quantum mechanics.  If quantum mechanics described in
Fig.~3 is combined with special relativity in Fig.~1, the result
will be the circle being squeezed into an ellipse.

FIG. 5.
Lorentz-squeezed space-time and momentum-energy wave functions.
As the hadron's speed approaches that of light, both wave function
become concentrated along their respective positive light-cone axes.
These light-cone concentrations lead to Feynman's parton picture.

FIG. 6.
Calculation of the parton distribution based on the harmonic
oscillator wave function.  It is possible to construct the covariant
harmonic oscillator wave functions for the three-quark system, and
compare the parton distribution function with experiment.  This graph
shows a good agreement between the oscillator-based theory and the
observed experimental data.


\newpage

\begin{thebibliography}{99}

\bibitem{bj69}
J. D. Bjorken and E. A. Paschos, Phys. Rev. {\bf 185}, 1975 (1969).

\bibitem{wig39}
E. P. Wigner, {\bf Ann. Math.} {\bf 40}, 149 (1939).

\bibitem{gribov66}
V. N. Gribov, B. L. Ioffe, and I. Ya. Pomeranchuk, J. Nucl. Phys.
(USSR) {\bf 2}, 768 or Sov. J. Nucl. Phys. {\bf 2}, 549 (1966).

\bibitem{ioffe69}
B. L. Ioffe, Phys. Lett. B {\bf 30}, 123 (1969).

\bibitem{drell71}
S. D. Drell and T. M. Yan, Ann. Phys. (NY) {\bf 60}, 578 (1971).

\bibitem{knp86} 
Y. S. Kim and M. E. Noz, {\em Theory and Applications of the Poincar\'e
Group} (Reidel, Dordrecht, 1986).

\bibitem{dir45}
For earlier efforts to construct covariant harmonic oscillators, see
P. A. M. Dirac, Proc. Roy. Soc. (London) {\bf A183}, 284 (1945);
H. Yukawa, Phys. Rev. {\bf 91}, 416 (1953);
M. Markov, Suppl. Nuovo Cimento {\bf 3}, 760 (1956);
V. L. Ginzburgh and V. I. Man'ko, Nucl. Phys. {\bf 74}, 577 (1965).

\bibitem{kno79a}
Y. S. Kim, M. E. Noz, and S. H. Oh, J. Math. Phys. {\bf 20}, 1341
(1979).

\bibitem{knp91} 
Y. S. Kim and M. E. Noz, {\it Phase Space Picture of Quantum Mechanics} 
(World Scientific, Singapore, 1991).

\bibitem{inonu53}
E. Inonu and E. P. Wigner, {\em Proc. Natl. Acad. Scie. (U.S.A.)}
{\bf 39}, 510 (1953).

\bibitem{barg48}
V. Bargmann and E. P. Wigner, {\em Proc. Natl. Acad. Scie (U.S.A.)}
{\bf 34}, 211 (1948).

\bibitem{hofsta55}
R. Hofstadter and R. W. McAllister, {\em Phys. Rev.} {\bf 98},
217 (1955).

\bibitem{gell64}
M. Gell-Mann, Phys. Lett. {\bf 13}, 598 (1964).

\bibitem{kim95mex}
For the latest review, see Y. S. Kim, {\it Wigner's Last Papers on
Spacetime Symmetries}, in the Proceedings of the Fourth International
Wigner Symposim, N. Atakisiev, T. Seligman, and K. B. Wolf, Eds.
(World Scientific, to be published).  See also
Y. S. Kim and E. P. Wigner, {\em J. Math. Phys.} {\bf 28}, 1175 (1987);
Y. S. Kim and E. P. Wigner, {\em J. Math. Phys.} {\bf 31}, 55 (1990).


\bibitem{fkr71}
R. P. Feynman, M. Kislinger, and F. Ravndal, {\em Phys. Rev.}
D {\bf 3}, 2706 (1971).

\bibitem{fey69}
R. P. Feynman, in {\em High Energy Collisions}, Proceedings of the Third
International Conference, Stony Brook, New York, C. N. Yang
{\em et al.}, eds. (Gordon and Breach, New York, 1969).


\bibitem{kim89} 
Y. S. Kim, Phys. Rev. Lett. {\bf 63}, 348-351 (1989).

\bibitem{dir49}
P. A. M. Dirac, Rev. Mod. Phys. {\bf 21}, 392 (1949).

\bibitem{kn73}
Y. S. Kim and M. E. Noz, Phys. Rev. D {\bf 8}, 3521 (1973);
D. Han and Y. S. Kim, Prog. Theor. Phys. {\bf 64}, 1852 (1980).

\bibitem{dir27}
P. A. M. Dirac, Proc. Roy. Soc. (London) {\bf A114}, 243 and 710 (1927).

\bibitem{fuji70}
K. Fujimura, T. Kobayashi, and M. Namiki, Prog. Theor. Phys. {\bf 43}, 
73 (1970).

\bibitem{hussar81}
P. E. Hussar, Phys. Rev. D {\bf 23}, 2781 (1981). 

\end{thebibliography}
\end{document}